\documentclass[reprint,aip,apl]{revtex4-1}

\usepackage[inline]{enumitem}

\usepackage{todonotes}                                   %for figure placeholders
\usepackage{outlines}

\usepackage{bm}                                              % for math
\usepackage{amssymb}   				% for math
\usepackage{mathtools} 			           % for math
\usepackage[version=4]{mhchem}                   %for chemical formula including \ce{}
\usepackage{natbib}
\begin{document}

\title{Vector-based categorization analysis with improved principal component generated reference}%
\author{Joseph M. Hamill}\email{joseph.hamill@dcb.unibe.ch} \affiliation{Department of Chemistry and Biochemistry, University of Bern, Freiestrasse 3, CH-3012, Bern, Switzerland}
\author{Matthias Arenz} \affiliation{Department of Chemistry and Biochemistry, University of Bern, Freiestrasse 3, CH-3012, Bern, Switzerland}
\date{\today}%

\begin{abstract}
With automated measurements, large data sets can be assembled with relative ease. Often these data sets have a large degree of variance. Nonetheless, we hope to find groups within the data set with shared distinguishing characteristics. Recently Albrecht, \textit{et al.} demonstrated a multi-parameter vector-based classification analysis which provided a powerful means to find clusters within a data set of conductance versus displacement traces measured from single molecular break junctions (Lemmer, \textit{et al., Nature Communications}, 2016, \textbf{7}, 12922). An idealized blank tunneling trace was used as a reference to calculate  three vector variables. The authors suggested a more appropriate reference can be chosen to better form clusters in the data. Here we propose using a principal component of the correlation matrix calculated from the data set to construct a fast and strategic reference trace. Principal components form an orthogonal basis set in units of the independent variables comprising the correlation matrix. The first principal component, the one corresponding to the largest eigenvalue of the correlation matrix, points in the direction of largest variance in the data. We used both a blank tunneling reference trace and a reference trace constructed from the first principal component of the correlation matrix to analyze a set of 4,4'-bipyridine single molecular break junctions. The blank tunneling reference trace produced a single elongated cluster in the vector analysis plot whereas the reference trace constructed from the principal component distinguished two arms in the cluster. We hypothesize that these two arms are due to groups of traces with either long or short molecular plateaus. Statistically based sorting algorithms, like vector-based categorization techniques, provide an objective framework to formulate such hypothesis and aid in designing new and pointed studies to verify them.
\end{abstract}

\maketitle

%--------------------------------------------------------------------------------------------------------------------%
%--------------------------------------------------------------------------------------------------------------------%
%\section{Introduction}
%--------------------------------------------------------------------------------------------------------------------%
%--------------------------------------------------------------------------------------------------------------------%
A principal goal of single molecular break junction (SMBJ) research is to develop models which predict from the structure of a molecule how it behaves electronically. Modern calculations provide a variety of powerful models to pinpoint the relationship between molecule structure and electron transfer through the molecule. Local currents\cite{Solomon2010}, quantum interference\cite{Markussen2010}, and side group moieties\cite{Higgins2017}, all effect electron transfer, and are all consequences of molecular structure. In recent years experimental techniques have also improved\cite{Meszaros2007} and a large number of break junctions can be measured reproducibly.\cite{Lemmer2016} However, there is still a large amount of variance within experiments. This variance is due to a number of different processes within the junction including electrode plasticity and molecular bonding geometry.\cite{Ulrich2006} For instance, it was shown that 4,4'-bipyridine achieved at least two metastable geometries within a single stretching process resulting in two plateaus in the conductance versus displacement trace.\cite{Perez-Jimenez2005JPCB,Quek2009NN} Reducing each experiment to a summary of averages misses these fine details. Experimental results seldom match calculations due in part to these averaged summaries. If large data sets can be separated into groups with shared features, it is likely that one or more of these groups will closely resemble the idealized geometries used in calculations.

Two different techniques were suggested for objectively sorting data sets in recent years. Here we show that when combined, the two strategies together provide a robust means to sort SMBJ data sets into groups.

Albrecht \textit{et al.} recently demonstrated an unsupervised multi-parameter vector-based classification procedure which sorted SMBJ traces into groups with shared features.\cite{Lemmer2016,Inkpen2015} The method calculated three variables from each trace and plotted the traces as points in cylindrical coordinates. If there existed groups in the data with strongly differentiable features, the different traces formed separate clusters in the cylindrical plot. The three variables were calculated by comparing each trace to a reference trace. Albrecht showed that a simple reference trace with a constant decay constant resembling a blank tunneling curve was sufficient to form clusters when the groups were sufficiently different. Furthermore, they showed that the results were not sensitive to the specific slope and offset of the tunneling curve employed. However, he also suggested that other reference traces may be used, for instance, a trace which optimized the variance in the data set. Such a reference trace would optimize the separation in the clusters.

Halbritter \textit{et al.} adapted the correlation matrix analysis common in spectroscopy studies to SMBJ data.\cite{Halbritter2010,Makk2012,Magyarkuti2017,Kurta2017} The approach calculates the correlation between 1D conductance histogram bins calculated from each trace in the data set. Halbritter showed that correlations between two conductances suggested there were plateaus in the traces at both conductances. Furthermore, anticorrelations suggest if there is a plateau at one conductance, there will not be a plateau in the correlated conductance. In the case of the two plateaus in 44BPY traces, Halbritter showed the conductances of those plateaus are anticorrelated because the shape of the two plateaus are different and contain, on average, different counts in the histogram.

The eigenvalues of any correlation matrix can be ordered from largest to smallest, and the corresponding eigenvectors make up an orthonormal basis set.\cite{Pearson1901} This is the basis for principal component analysis (PCA).\cite{hotelling1933analysis,Shlens2014,Bro2014} The first eigenvector, called the first principal component ($PC_1$), points in the direction of largest variance in the data set and all subsequent PCs point in ever-decreasing orthogonal directions of variance. Thus by diagonalizing a data set's correlation matrix we determine a valuable set of reference traces for use in a vector-based classification procedure.

In previous work we showed\cite{Hamill2017} the first and second principal components of the correlation matrix introduced by Halbritter was an effective means to sort data sets into groups with shared features. In those studies the correlation matrix was calculated from 1D conductance histograms from each trace in the data set. For the present study this will not suffice. To use the vector-based categorization approach suggested by Albrecht, the conductance versus displacement trace itself was used for further analysis.

The goal of this study was to show that by using the first principal component to construct a reference trace to compare experimental traces against, the clusters in the vector analysis plot are better spread into groups for further separation and focused study. We first applied an analysis approach similar to Albrecht, \textit{et al.}, using an idealized tunneling curve as a reference trace. This approach produced a single elongated cluster on the vector analysis plot. We then analyzed the same data with a reference trace constructed from the first principal component of the data matrix constructed from the plateau region of the conductance versus displacement traces. This improved reference trace produced a cluster with two elongated arms.

\begin{figure}
\includegraphics[width=8.3cm]{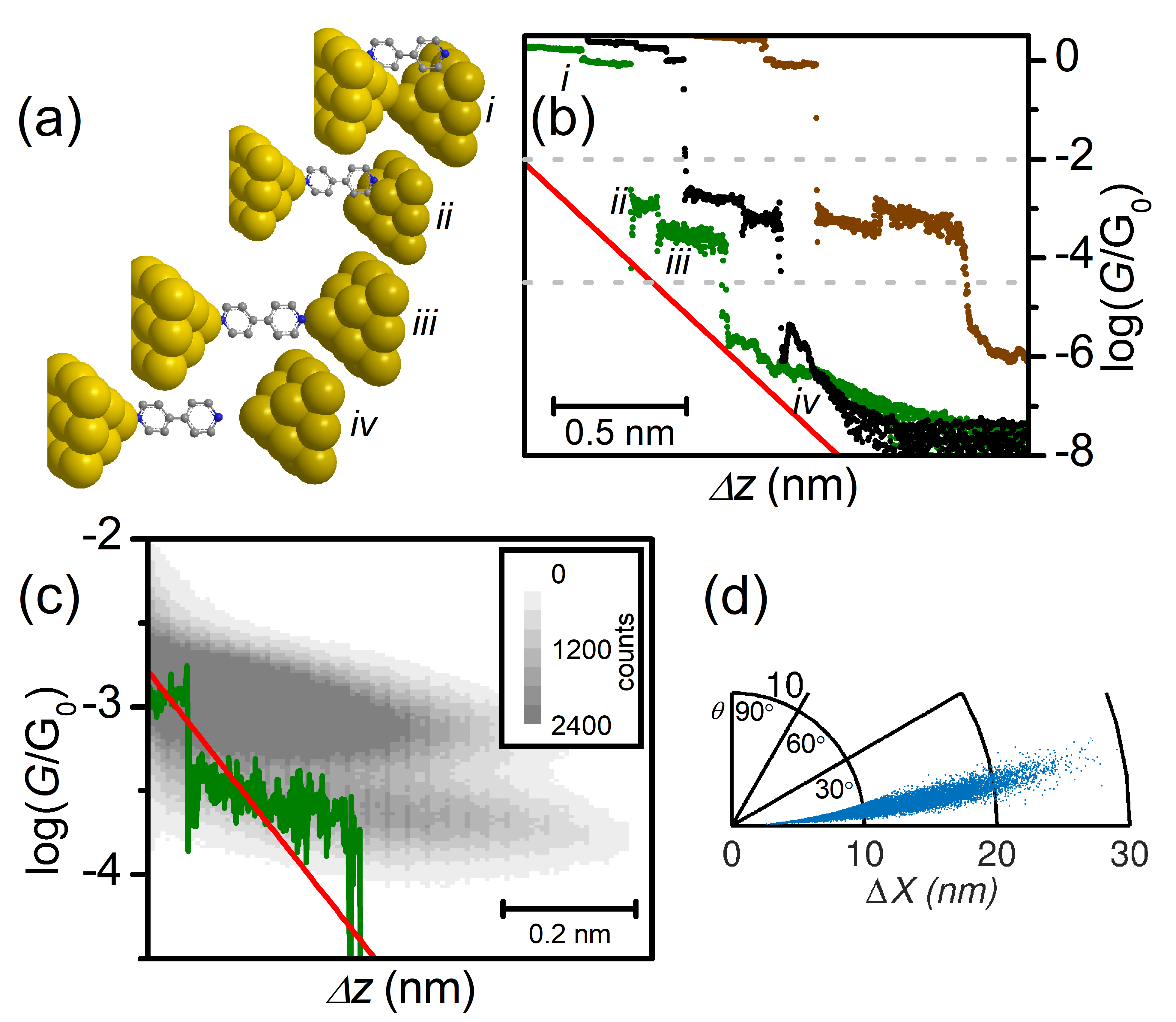}
\caption
{\label{fig:one} Experimental results of 26000 MCBJ traces with 44BPY in solvent.  
(a) 44BPY break junction progression from closed gold-gold contact (\textit{i}), first metastable 44BPY geometry (\textit{ii}), fully extended 44BPY geometry (\textit{iii}), and open junction (\textit{iv});
(b) tunneling reference trace (red) and example traces; gray dashed lines mark plateau region;
(c) 2D conductance versus displacement histogram of plateau region with example trace (green) and tunneling reference curve (red);
(d) $\Delta X$ versus $\theta$ for each trace calculated against the tunneling reference curve plotted in polar coordinates.
}
\end{figure}

%--------------------------------------------------------------------------------------------------------------------%
%--------------------------------------------------------------------------------------------------------------------%
%\section{Results and discussion}
%--------------------------------------------------------------------------------------------------------------------%
Conductance versus electrode tip displacement traces were measured using mechanically controlled break junctions (MCBJ).\cite{Huang2015break} MCBJs work by fixing a notched gold wire (0.1 mm diameter, Goodfellow, 99.99\%) on a steel spring sheet substrate. A piezoelectric rod arches the substrate and breaks the gold wire open at the notch. The break junction was immersed in a solution of $0.1~\textrm{mM}$ 4,4'-bipyridine (44BPY, Sigma-Aldrich, 98\% with no further purification) in 1:4 tetrahydrafuran:mesitylene organic solvent (THF: Sigma-Aldrich, with 250 ppm BHT inhibitor, 99.9\% with no further purification; TMB: Sigma-Aldrich, 98\% with no further purification). The conductance of the junction was measured while the junction broke open. The trace was plotted on a semilog plot of conductance (scaled by the conductance quantum, $\textrm{G}_0$\footnote{the conductance quantum is the conductance of a single 1D conducting channel, $2e^2/h$, where $e$ is the charge on the electron and $h$ is Planck's constant.}) versus displacement of the electrodes (example traces can be seen in Fig.~\ref{fig:one}(b) in green, gray and brown). When the junction was closed, the conductance was saturated above $10^0~\textrm{G}_0$. Immediately before it broke open, an atomic gold-gold junction was formed, and the conductance was exactly $10^0~\textrm{G}_0$ [cartoon \textit{i} in Fig.~\ref{fig:one}(a)]. After the gold-gold junction broke, the 44BPY binded between the two electrodes, and a plateau was seen in the conductance trace at the single molecular conductance  [cartoons \textit{ii} and \textit{iii} in Fig.~\ref{fig:one}(a)]. Finally, the gold-44BPY bond broke and the conductance dropped to the open circuit sensitivity of the device at about $10^{-7}~\textrm{G}_0$ [cartoon \textit{iv} in Fig.~\ref{fig:one}(a)]. Break junctions were repeated at a rate of approximately 3 Hz until over 26000 traces were measured.

All traces were binned into a 2D conductance versus displacement histogram [Fig.~\ref{fig:one}(c)]. The two clouds in Fig.~\ref{fig:one}(c) were caused by two consecutive geometries of 44BPY in the junction as it elongates. The first geometry was a tilted orientation bringing the $\pi$ orbitals of one pyridine in close proximity to one gold electrode and the second geometry was a perpendicular orientation [cartoons \textit{ii} and \textit{iii} in Fig.~\ref{fig:one}(a)] The green example trace exhibits two plateaus corresponding to these two clouds.

The data was analyzed using the vector analysis method described in Ref.~\cite{Lemmer2016}. The vector analysis method required a reference trace to compare each trace against. First, as suggested by the authors of this study, we used an artificially produced perfect tunneling curve beginning at $I_0 = 20~\textrm{nA}$ and with a tunneling decay constant of $\beta = 0.5 \textrm{\AA}^{-1}$ [red curve in Figs.~\ref{fig:one}(b) and (c)]. 

The vector distance, $\Delta X_m$ between each trace, $\mathbf{X}_m$, and the reference trace, $\mathbf{R}$, was calculated from
\begin{equation}\label{eq:delta}
 \Delta X_m = |\mathbf{X}_m - \mathbf{R}|
\end{equation}
and the vector angle, $\theta_m$, between each trace and the reference trace was calculated from
\begin{equation}\label{eq:theta}
 \theta_m = - \frac{\mathbf{R}\cdot\mathbf{X_m}}{|\mathbf{R}|*|\mathbf{X_m}|}.
\end{equation}

$\Delta X$ and $\theta$ were calculated against the red reference curve and each trace in the data set was plotted on a polar scatter plot [Fig.~\ref{fig:one}(d)]. This procedure resulted in a single elongated cluster which suggested a large degree of uniformity in the data set. 

The blank tunneling curve provided a simple, reproducible option for a reference trace. Albrecht showed that $I_0$ and $\beta$ can be varied with little effect on the interpretation of the results. However, in this case the tunneling reference trace was not sufficient to separate clusters in this data set. 

\begin{figure}
\includegraphics[width=8.3cm]{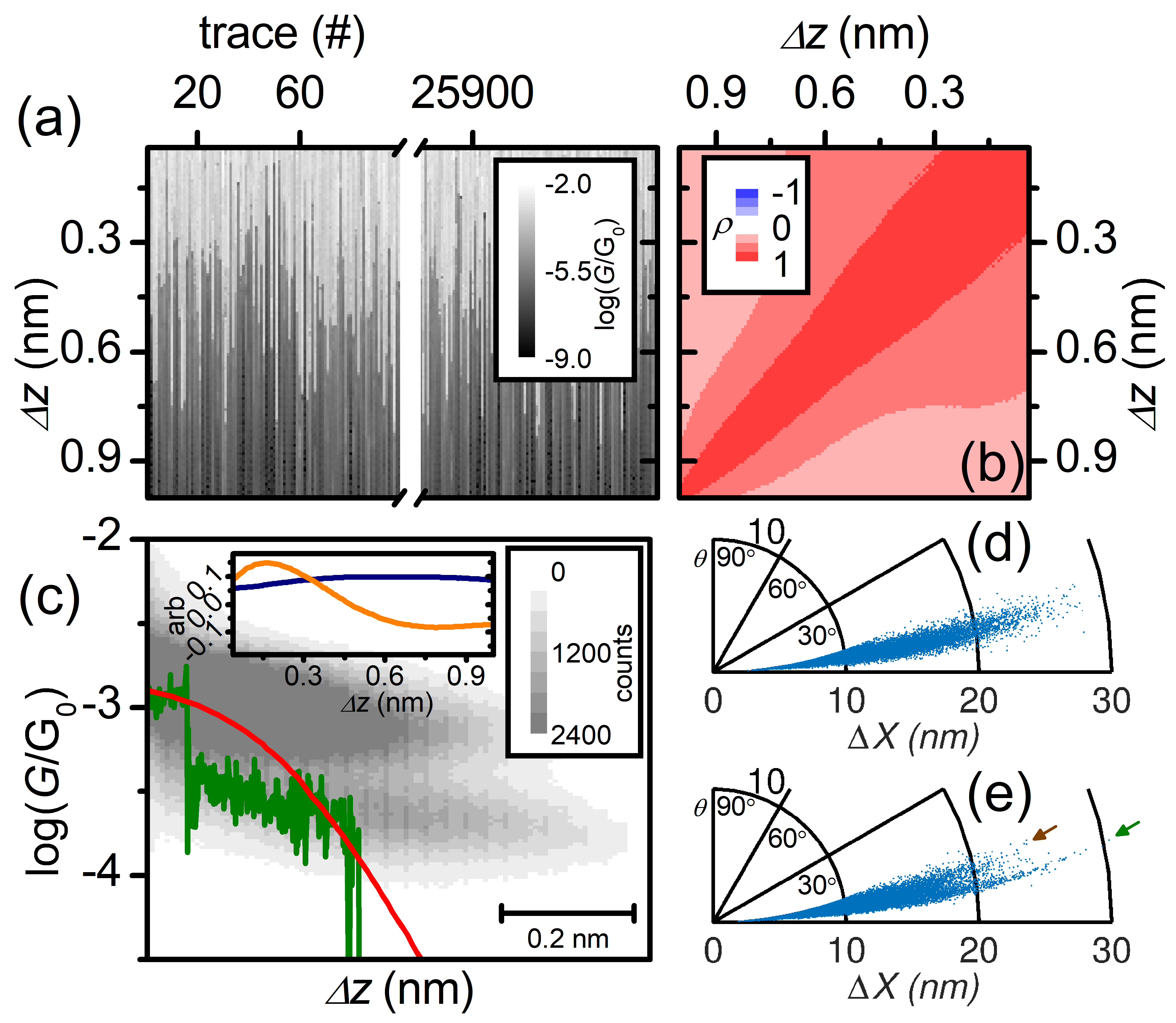}
\caption
{\label{fig:two} Vector analysis using principal component as reference curve.  
(a) Data matrix with approximately 26000 conductance versus displacement plateau traces interpolated to 128 displacements as columns;
(b) 2D correlation intensity plot calculated from (a);
(c) 2D conductance versus displacement histogram with example trace (green) and principal component PC  reference trace (red) with mean added back; inset is $PC_1$ and $PC_2$ calculated from (b);
(d) vector analysis using PC as reference trace showing two arms (maroon and green arrows.
}
\end{figure}

Applying the same method with a more carefully chosen reference trace was tried next. The new reference trace was constructed from the first principal component of the correlation matrix of the data set. For this, the data matrix must comprise a column for each trace, with each column containing a measurement at set displacements over the range of the molecular plateaus ($0-1~\textrm{nm}$) [Fig.~\ref{fig:two}(a)]. To achieve this, the segment of each trace in this range was selected and interpolated to 128 defined displacements. The conductance trace along these interpolated segments formed the data matrix. From this data matrix a correlation matrix was calculated [Fig.~\ref{fig:two}(b)]. The correlation matrix described variables (different displacements) which were correlated with one another. In the case of the 44BPY data all displacements were correlated because the trend of each trace was to decrease in conductance as the displacement increased. Because the correlation matrix was rather featureless, the first and second principal components [$PC_1$ and $PC_2$, Fig.~\ref{fig:two}(c) inset] were smooth. $PC_1$ described an almost constant dependence of each displacement with the others, with a slight increase as the displacement increased.

For the purpose of demonstrating a straightforward approach, $PC_1$ was used to create the reference trace in the following analysis. The first principal component of a correlation matrix describes the direction of largest variance in the data set. 
$PC_1$ therefore provided the best means to observe separable clusters in the vector analysis. Even if the tunneling reference curve was not able to distinguish separate clusters in the data, a reference trace constructed from the first principal component is guaranteed to distinguish variances in the data, if they exist, so that they form separate clusters.

$PC_1$ described the direction of the most variance in the data set with the mean removed from each variable. This was a consequence of calculating the correlation matrix. To be used as a reference trace, the mean needed to be re-added to $PC_1$. The new reference trace derived from $PC_1$ and describing the direction of largest variance was the red curve in Fig.~\ref{fig:two}(c). Calculating $\Delta X$ and $\theta$ using Eqs.~\ref{eq:delta} and \ref{eq:theta} and the new reference trace yielded Fig.~\ref{fig:two}(d). The single elongated cluster from the tunneling reference curve was branched into two arms when the principal component reference curve was used [red and green arrows in Fig.~\ref{fig:two}(d)].

The distribution was analyzed using density-based spatial clustering of applications with noise (DBSCAN\cite{ester1996density}) as a quick and accessible clustering algorithm available in Octave.\cite{Octave} For a range of neighborhood sizes and densities, the algorithm always returned a cluster count of one. Thus we cannot conclude that the two arms are different clusters. Nonetheless, their bimodal distribution will in the future allow for the two arms to be separated for further analysis. Each arm describes different and distinguishable behaviors within the break junction experiment, and each arm should provide new insight to further inform new directions for calculations and experimental designs.

Figure~\ref{fig:two}(c) showed $PC_1$  became slightly larger as the plateau was longer. Because of this, we hypothesize the reference trace constructed from $PC_1$ distinguished between shorter and longer molecular plateaus. Shorter traces, which more closely resembled the reference trace, formed points in the arm with the green arrow in Fig.~\ref{fig:two}(e), where $\theta$ was smaller. Longer traces, which resembled the reference trace less, formed points in the arm with the maroon arrow in Fig.~\ref{fig:two}(e). This hypothesis needs further analysis to confirm.

%--------------------------------------------------------------------------------------------------------------------%
%--------------------------------------------------------------------------------------------------------------------%
%\section{Conclusion}
%--------------------------------------------------------------------------------------------------------------------%
%--------------------------------------------------------------------------------------------------------------------%
Here we demonstrated the use of principal components as a simple strategic approach to determining an optimized reference trace in a multi-parameter vector-based classification analysis. Using a simple tunneling trace as reference failed to form clusters in a large data set of molecular break junction traces. Substituting a reference trace derived from the first principal component of the data succeeded in distinguishing a bimodal distribution where the tunneling reference trace only distinguished a single mode distribution. The principal component was calculated from the correlation matrix using 128 displacement points along the molecular plateau as variables and conductance as the observable. Thus, determining the principal component required only four more operations on the data set. First, interpolating the conductance trace to a defined number of points along the plateau. Second, calculating the correlation matrix. This is a standard procedure in SMBJ research. Third, calculating the first principal component of the correlation matrix. Finally, constructing the reference trace.

Multi-parameter vector-based classification techniques provide a way to sort data into groups so that the groups can be studied and compared to calculations separately. Furthermore, they can provide a way to quantitatively determine how distinguishable the clusters are, and thus how well the groups are separated. The next steps in our above analysis is to analyze the arms formed using the PC reference trace using alternative clustering algorithms which may separate the arms into two or more clusters.  If this is not successful, another statistical means may be derived to objectively separate the two arms into two groups for further analysis. Studying the two arms separately will reveal new information about the distribution of traces within a 44BPY experiment. The strength of statistically based sorting algorithms is that any groups are determined objectively and therefore point to significant and reproducible results. For instance, our hypothesis that 44BPY traces can be grouped into traces with either long or short molecular plateaus. In the future these arms can be studied separately. For instance, we hypothesized the distinction between the two arms of the vector plot was due to groups of traces with either long or short plateaus. If further analysis on this system proves this hypothesis to be accurate, then further experimental and theoretical studies can be pointedly designed to determine the cause of these different lengths and why they seem to be discrete, i.e. either long or short, but not all the lengths in between.

Finally, this new approach can be applied to other molecular break junction systems, and also to other stochastically distributed data sets from other fields. Any variables may be included in vector-based categorization analysis, as well as in principal component analysis. The approach demonstrated here may be applied broadly to any and all variables.

%--------------------------------------------------------------------------------------------------------------------%
%--------------------------------------------------------------------------------------------------------------------%
%\section*{Acknowledgments}
%--------------------------------------------------------------------------------------------------------------------%
%--------------------------------------------------------------------------------------------------------------------%
\begin{acknowledgments}
This work was generously supported by the EC FP7 ITN ``MOLESCO'' project number 606728, and the University of Bern.
\end{acknowledgments}

%--------------------------------------------------------------------------------------------------------------------%
%--------------------------------------------------------------------------------------------------------------------%
%Bibliography
%--------------------------------------------------------------------------------------------------------------------%
%--------------------------------------------------------------------------------------------------------------------%

%\printbibliography
%\bibliography{G:/Dropbox/hide/References/Jabref/library}

%

\end{document}